\documentclass{article}

\usepackage[final]{nips_2017}

\usepackage[utf8]{inputenc} 
\usepackage[T1]{fontenc}    
\usepackage{hyperref}       
\usepackage{url}            
\usepackage{booktabs}       
\usepackage{amsfonts}       
\usepackage{nicefrac}       
\usepackage{microtype}      
\usepackage{graphicx}

\title{Exploring large scale public medical image datasets}

\author{
  Luke Oakden-Rayner \\
  The Australian Institute for Machine Learning\\
  The School of Public Health, The University of Adelaide\\
  The Royal Adelaide Hospital\\
  \texttt{luke.oakden-rayner@adelaide.edu.au}\\
  \And
}

\begin{document}

\maketitle

\begin{abstract}
Rationale and Objectives: Medical artificial intelligence systems are dependent on well characterised large scale datasets. Recently released public datasets have been of great interest to the field, but pose specific challenges due to the disconnect they cause between data generation and data usage, potentially limiting the utility of these datasets.

Materials and Methods: We visually explore two large public datasets, to determine how accurate the provided labels are and whether other subtle problems exist. The ChestXray14 dataset contains 112,120 frontal chest films, and the MURA dataset contains 40,561 upper limb radiographs. A subset of around 700 images from both datasets was reviewed by a board-certified radiologist, and the quality of the original labels was determined.

Results: The ChestXray14 labels did not accurately reflect the visual content of the images, with positive predictive values mostly between 10\% and 30\% lower than the values presented in the original documentation. There were other significant problems, with examples of hidden stratification and label disambiguation failure. The MURA labels were more accurate, but the original normal/abnormal labels were inaccurate for the subset of cases with degenerative joint disease, with a sensitivity of 60\% and a specificity of 82\%.

Conclusion: Visual inspection of images is a necessary component of understanding large image datasets. We recommend that teams producing public datasets should perform this important quality control procedure and include a thorough description of their findings, along with an explanation of the data generating procedures and labelling rules, in the documentation for their datasets.

\end{abstract}

\section{Introduction}

The successful training of modern artificial intelligence (AI) relies on large, well-characterised datasets\textsuperscript{1}. The availability of these datasets can be considered a major barrier to the production of high quality image analysis AI systems in radiology, not only because the cost to produce these datasets is high, but also because access to existing datasets is restricted. Privacy concerns around the sharing of medical data and the competitive advantage that medical AI companies obtain from their own proprietary datasets has severely limited the sharing of these resources.

To overcome this challenge, several large public datasets have been made available in recent years. The ChestXray14 (CXR14) dataset produced by a team of researchers at the National Institutes of Health Clinical Center contains over 112,000 chest radiographs\textsuperscript{2}. The Musculoskeletal Radiology (MURA) dataset and competition from the Stanford Machine Learning Group contains over 40,000 upper limb radiographs\textsuperscript{3}. The Radiological Society of North America (RSNA) Paediatric Bone Age challenge dataset contains 14,236 upper limb radiographs\textsuperscript{4}. Several other notable recent releases of data include The RSNA pneumonia challenge, which builds on the CXR14 dataset with radiologist-produced labels, the QC500 dataset from Qure.AI (a commercial group) containing 500 CT head images in patients with and without intracranial haemorrhage\textsuperscript{5}, and the fastMRI dataset from New York University and Facebook AI Research containing 10,000 knee MRI studies\textsuperscript{6}. Each of these datasets, other than the fastMRI dataset, are accompanied by labels (indicators of a particular disease or imaging finding within each study) that are intended to inform the training of clinically useful AI systems.

These public datasets have generated an enormous level of interest in the medical image analysis community. 260 teams registering for the RSNA Bone Age challenge, and 346 teams registered for the RSNA Pneumonia detection challenge. Similarly, dozens of teams have published results on the CXR14 and MURA datasets.

This democratisation of access to large-scale medical data has undoubtedly been of benefit to the medical image analysis community, however it is important to understand the specific challenges presented by these public datasets.

The root of the problem with public datasets is that the development processes (data gathering, cleaning, and labelling) are disconnected from the usage of the data. This means that the end-user of the data may not understand the nuances of the development processes, including many subtle design decisions that are not always well communicated in the published reports on the datasets. This problem is compounded by the highly opaque nature of medical images to non-experts (i.e., radiologists). Unlike with datasets of ordered rows and columns of numbers, where the relationships between input variables and labels can be analysed by anyone practiced in the skills of data science, medical imaging datasets require a high level of expertise to understand.

This disconnect between dataset development and usage can cause a variety of problems; 1) the accuracy of the labels can be overestimated by users, particularly when the weaknesses of the label generation procedures are poorly explained, 2) the presence of unlabelled visual subsets (hidden stratification) can significantly alter the usefulness of the labels in training AI systems, and 3) the clinical meaning of the labels themselves can be obscure. Making matters worse, if AI systems are then tested on data generated with the same procedures (i.e., on test data drawn from the same dataset), then these problems may occur silently; the results of testing can look good because the models can learn to reproduce the flawed labels from the training data, but the actual clinical performance of these systems will be poor. 

Each of these problems can only be overcome by the direct application of medical knowledge; an expert must appreciate the presence of subsets, review the quality of the labels, and comprehend the logic of the label schema (the rules that govern what each label means, and their relationships to each other). Only after this evaluation is it possible to determine the value of the dataset for building medical AI systems.
In this work we explore two large public datasets to demonstrate the importance of this review process, assessing the accuracy of the provided labels, as well identifying other issues that may limit the utility of these datasets. In doing so, we stress the importance of expert visual analysis as a form of quality control when building and using these large scale datasets, and present recommendations for teams planning to release public datasets in the future.

\section{Methods}
\subsection{Datasets}
\subsubsection{CXR14}
The CXR14 dataset is a large-scale dataset for pathology detection in chest radiographs. This dataset was released in 2017 and updated later the same year, containing 112,120 frontal chest films from 30,805 unique patients. The dataset is drawn from a single tertiary medical centre (the NIH Clinical Center) and appears to include films from multiple clinical settings, including intensive care unit (ICU) and non-ICU patients.

The images had a resolution of 3000 x 2000 pixels, and were in the DICOM format (which stores greyscale pixels with around 3000 to 4000 grey levels). These were downsampled into PNG images with a resolution of 1024 x 1024 pixels and 255 grey levels.

The dataset was labelled using natural language processing on the original (clinical) free-text reports, a process that involved matching keywords related to various forms of pathology, and identifying negations (sentences that exclude certain findings and pathologies).

The dataset was initially labelled with 8 different classes, however this was expanded to 14 classes later in 2017. These classes were: atelectasis, consolidation, infiltration, pneumonia, cardiomegaly, pneumothorax, fibrosis, pleural effusion, mass, nodule, pleural thickening, oedema, hiatus hernia, emphysema, and a normal (no finding) class. As the dataset was collected from a clinical archive, these image findings occur roughly at clinical prevalence, ranging from less than 0.5\% (hernia) up to almost 10\% (infiltration). The “normal” or “no finding” class makes up around 75\% of the total images, or roughly 84,000 studies.

Notably, there are many patients with multiple x-rays. Almost half of all patients (13,302) had more than one study, together accounting for 84\% of the data. Further exploration of the extent of multiple studies is provided in Table 1, revealing that the number of patients with numerous repeat images account for a surprisingly large proportion of the dataset. In many of these cases, these will reflect ICU patients who have repeat imaging daily, where the images change very little across the entire series, significantly reducing the diversity of the dataset.

\begin{table}[t]
  \centering
  \begin{tabular}{lllllll}
    \toprule
    \cmidrule{1-3}
    Number of studies per patient & Number of patients & Total number of studies \\
      &  (percentage of patients  &  (percentage of studies \\
    & in CXR14) & in CXR14) \\
    \midrule
    > 1 & 13,302 (43\%) & 94,617 (84\%) \\
    > 5 & 4,821 (16\%) & 70,081 (63\%) \\
    > 10  & 2,225 (7\%)	& 50,468 (45\%) \\
    > 50  & 151 (0.5\%) &	10,812 (10\%) \\
    > 100  & 18 (<0.001\%) & 2,310 (2\%) \\
    \bottomrule
  \end{tabular}
  \\

\begin{flushleft}
Table 1:  The prevalence of patients with multiple studies in the CXR14 dataset. 
\end{flushleft}

\end{table}

\subsubsection{MURA}
The MURA dataset is a large dataset for abnormality detection in upper limb musculoskeletal radiographs. Released in 2018, the dataset contains 40,561 images from 14,863 studies, obtained from a single tertiary medical centre (Stanford Hospital). The dataset includes seven standard upper limb study types, with studies of the fingers (2110 studies), hands (2185 studies), wrists (3697 studies), forearms (1010 studies), elbows (1912 studies), humeri (727 studies), and shoulders (3015 studies).

The dataset was labelled at the time of clinical interpretation by board-certified radiologists, each providing a label of “normal” or “abnormal” at the time of performing their usual report. No additional pathology specific labels were produced.

It was not specified in the MURA paper what constituted an “abnormal” finding, however an analysis of the label composition was performed by the authors; the text reports of 100 cases labelled abnormal were reviewed and the specific abnormalities identified were noted. This analysis revealed, of 100 abnormal cases, there were 53 studies with fractures, 48 studies with implanted hardware (such as joint replacements), 35 studies with degenerative joint disease, and 29 studies with other abnormalities such as lesions and subluxations.

The images in MURA was also downsampled, from DICOM images with a resolution of 1500 x 2000 pixels and around 3000 to 4000 grey levels, to PNG images with a resolution of 512 x 200-400 pixels, and 255 grey levels.

Multiple views were performed for most cases, with over 92\% of cases having more than 1 image. There were relatively few repeat studies however, with only around 4\% of patients having 2 or more studies.

\subsection{Visual inspection}
The accuracy of the datasets was assessed by visual inspection performed by LOR, a board-certified radiologist. Each dataset was relabelled by LOR according to the findings of an initial exploratory assessment. Visual review of a randomly generated subset of around 100 images from each class (i.e., pneumonia, cardiomegaly, etc. in CXR14, and normal, abnormal in MURA) was performed. This exploration was to understand the images and labels, and to identify any common problems with the label schemata, with the findings used to inform the relabelling process. The images used for exploratory analysis were separate from the images that were used for relabelling.

\subsubsection{CXR14}

In the CXR14 dataset, large-scale relabelling at the original prevalence was not achievable due to the rarity of many of the labels present in the original dataset and the ambiguity of many of the label classes. For example the CXR14 label schema considers pneumonia, consolidation, and infiltration as distinct processes, but clinically it is rarely possible to distinguish these processes, at least without clinical information. As such, any labels created for these categories would likely be idiosyncratic and unfairly reflect upon the accuracy of the original labels.

To overcome these issues, an enriched, randomly selected subset of 50 cases per class were reviewed, for a total of 700 cases. Rather than attempting to relabel each case with the 14 possible classes, each case was reviewed purely for the presence of the label(s) it had been given; for example, a case from the cardiomegaly subset was re-labelled “cardiomegaly” or “not cardiomegaly”. 

As each label class is not explicitly defined in the original paper, a permissive labelling rule was applied. In general, if a case could plausibly reflect the finding in the original label, it was considered positive for that finding, as long as that finding was visible to the eye of faith (i.e., an ill-defined basal opacity could be positive for pleural effusion or consolidation, but not for a mass). This is in comparison to labelling rules such as “label only findings that you would report in clinical practice” or “label only findings that you are certain of”, both of which rules are much stricter than the rule applied. As such, the labels should diverge as little as possible from the original CXR14 labels, while still reflecting the visual appearance.
Notably this label rule was more permissive than the rules used in previous analysis of this dataset\textsuperscript{7}.
Because this approach was very permissive, a second rule which was closer to normal clinical practice was also applied for relabelling. The finer details of both of these rules are provided in the Supplement.

\subsubsection{MURA}

In MURA, relabelling of 714 randomly selected cases was performed. This was achievable at the original prevalence of the dataset, as the ratio of abnormal to normal cases was around 45:55. 

Pathology specific labelling was undertaken, with labels produced to identify cases containing fractures, implanted hardware, degenerative joint disease, bone tumours, and a class containing miscellaneous pathologies (such as osteopaenia, subluxations, and ligamentous injuries). These labels were produced without knowing the original MURA label (normal or abnormal) for each study.

The labelling rule in this case was less permissive, as these labels were part of a larger effort to relabel the MURA dataset with clinically accurate labels. As such, cases were labelled to the best accuracy of the radiologist.

\subsection{Analysis}
\subsubsection{CXR14}
The CXR14 data labels were assessed by calculating their positive predictive value (PPV), using the expert visual labels as the gold-standard. The PPV was presented here because the cases were only assessed for the presence or absence of their associated labels findings. As explained earlier, the dataset could not be efficiently relabelled wholesale for a variety of reasons, which meant that the negative predictive value could not be determined. As such it should not be assumed that these results reflect a comprehensive assessment of the dataset, but instead simply provide evidence towards the quality of the labels.

\subsubsection{MURA}
The MURA data labels were assessed using the expert visual labels as the ground truth, with the sensitivity and specificity calculated. In this dataset, comprehensive relabelling was possible, which allowed for assessment of both positive and negative cases.

In both datasets, subgroup analysis was performed. In the MURA dataset, the subgroups included the body region imaged and the specific pathology groups labelled by visual review. In the CXR14 dataset, the specific subgroups beyond just the label classes were identified during exploratory analysis of the images.

\section{Results}
\subsection{CXR14}
50 cases from each of the 15 class groups were assessed by LOR. The results for the visual assessment of the CXR14 dataset are provided in Table 2. Even with the use of permissive labelling rules, the PPV determined by visual assessment of the images is below the estimated PPV presented in Wang et al. in all classes.

\begin{table}[t]
  \centering
  \begin{tabular}{llll}
    \toprule
    \cmidrule{1-4}
    Disease class & PPV (visual, permissive) & PPV (visual, clinical) & PPV (text mining, \\
     & & &  from Wang et al.$^{\dag}$) \\
    \midrule
    Consolidation	& 80\% & 66\%	& - \\
	Atelectasis	& 80\% & 50\%	& 99\% \\
	Infiltration	& 66\% & 36\%	& 74\% \\
	Pneumonia	& 60\% & 50\%	& 66\% \\
	Oedema	& 76\% &     40\%	& - \\
	Nodule	& 76\% & 	64\%	& 96\% \\
	Mass	& 64\%	& 46\%	& 75\% \\
	Pneumothorax$^{\alpha}$	& 90\% (60\%)	& 90\% (60\%)	& 90\% \\
	Pleural effusion	& 74\%	& 70\%	& 93\% \\
	Pleural thickening	& 84\%	& 52\%	& - \\
	Emphysema	& 14\%	& 10\%	& - \\
	Cardiomegaly	 & 70\%	& 52\%	& 100\% \\
	Fibrosis	& 46\%	& 26\%	& - \\
	Hernia$^{\beta}$	& 94\%	& 78\%	& - \\
	Normal	& 76\%	& 62\%	& 87\% \\
    \bottomrule
  \end{tabular}
  \\

\begin{flushleft}
Table 2:  Visual assessment of the CXR14 dataset labels, using both permissive and “clinical-style” relabelling rules as the ground-truth. \\
$^{\alpha}$The pneumothorax class was stratified, with the majority of images containing chest drains. The PPV of the subset of cases without chest drains is given in parentheses. \\
$^{\beta}$The hernia class was almost always correctly labelled, but on exploratory analysis there were many examples of false negatives that were not captured in the PPV value. \\
$^{\dag}$The text mining PPV reported by Wang et al. was scored against the Open-i dataset9, rather than using data from the  CXR14 cohort. The labels with no text mining PPV did not appear in the Open-i dataset.The prevalence of patients with multiple studies in the CXR14 dataset. 
\end{flushleft}

\end{table}

Exploratory visual analysis revealed two striking examples of visual stratification. The first is in the pneumothorax class, where 80\% of the positive cases have chest drains. In these examples, there were often no other features of pneumothorax (i.e., the lung did not appear collapsed, likely reflecting a successfully treated pneumothorax). While the overall PPV was quite high (90\%), of the cases without chest drains the PPV was lower, at 60\%.

The second example of visual stratification was related to the emphysema class. The majority of cases (86\%) had subcutaneous emphysema rather than pulmonary emphysema. This is almost certainly a specific failure of the original labelling method, where these keywords were not successfully disambiguated. This resulted in a very low PPV for the emphysema labels.

\subsection{MURA}
The imaging characteristics of the test subset, as well as the entire MURA dataset, are provided in Table 3. The distribution of cases within these groups was similar.

\begin{table}[t]
  \centering
  \begin{tabular}{lllll}
    \toprule
    \cmidrule{1-5}
    & No. in test subset & Percentage of the  & No. in MURA dataset & Percentage of   \\
    & (\% of the subset)	 & test subset	& (\% of the MURA dataset) & MURA labelled  \\
    & & labelled abnormal & & abnormal \\
    \midrule
	All images	& 714 (100\%)	& 43\%	& 14656 (100\%)	& 39\% \\
	Finger	& 126 (18\%)	& 44\%	& 2110 (14\%)	& 35\% \\
	Hand	& 131 (18\%)	& 44\%	& 2185 (15\%)	& 27\% \\
	Wrist	& 182 (25\%)	& 36\%	& 3697 (25\%)	& 38\% \\
	Forearm	& 45 (6\%)	& 28\%	& 1010 (7\%)	& 35\% \\
	Elbow	& 70 (10\%)	& 33\%	& 1912 (13\%)	& 38\% \\
	Humerus	& 19 (3\%)	& 42\%	& 727 (5\%)	& 47\% \\
	Shoulder	& 141 (20\%)	& 52\%	& 3015 (21\%)	& 52\% \\
    \bottomrule
  \end{tabular}
  \\

\begin{flushleft}
Table 3: The imaging characteristics of the relabelled test subset, and the MURA dataset overall.
\end{flushleft}

\end{table}

Taking the expert review of the images as the ground-truth, the sensitivity (true positive rate) and specificity (true negative rate) of the MURA labels is presented in Table 4, both overall and by region.

\begin{table}[t]
  \centering
  \begin{tabular}{lll}
    \toprule
    \cmidrule{1-3}
    & Sensitivity & Specificity \\
    \midrule
	All images	& 80\%	& 75\% \\
	Finger &	72\%	& 82\% \\
	Hand	& 79\%	& 80\% \\
	Wrist	& 63\%	& 89\% \\
	Forearm	& 56\%	& 90\% \\
	Elbow	& 94\%	& 81\% \\
	Humerus	& 100\%	& 92\% \\
	Shoulder	& 82\%	& 64\% \\
    \bottomrule
  \end{tabular}
  \\

\begin{flushleft}
Table 4: The sensitivity and specificity of the MURA labels, overall and by region. Specific subgroups where the labels underperform compared to the average performance across the dataset are highlighted in bold.
\end{flushleft}

\end{table}

The class specific (i.e., per pathology) sensitivity of the MURA labels is presented in Table 5, using the expert class labels as the ground truth. The specificity of the MURA labels is also presented, but the values are inflated by the low per-class prevalence of the conditions relative to the number of normal studies.

\begin{table}[t]
  \centering
  \begin{tabular}{lll}
    \toprule
    \cmidrule{1-3}
    & Sensitivity & Specificity \\
    \midrule
	Fracture	& 92\%	& 98\% \\
	Hardware &	85\%	& 98\% \\
	Degenerative disease	& 60\%	& 82\% \\
	Other	& 82\%	& 97\% \\
    \bottomrule
  \end{tabular}
  \\

\begin{flushleft}
Table 5: The sensitivity and specificity of the MURA labels by pathology type. The specificity values appear high partially due to the low per-class prevalence of the findings.
\end{flushleft}

\end{table}

There was poor identification of degenerative joint disease by the MURA labels (sensitivity = 60\%) compared to fractures and hardware (sensitivity = 92\% and 85\% respectively), which was also reflected in lower sensitivity for identifying pathology in the regions typically affected by joint disease (i.e., in the wrist and the forearm, but not in the elbow or humerus). 

During this analysis, it was noted that there was an unexpected number of false positive results in the shoulder class; that is, cases that were labelled abnormal in MURA but were considered normal on visual review. Visual exploration of these cases revealed no clear patterns. There were several missed diagnoses amongst the visually reviewed cases (several subtle fractures and two lytic bone lesions), but the majority of the false positive cases revealed no identifiable pathology.

\section{Discussion}
The two datasets explored were of variable quality. The PPV of the labels in the CXR14 dataset were typically quite low, even allowing for differences in reporting style and inter-observer variability. By contrast, the MURA labels were of much higher accuracy, other than in the subset of patients with features of degenerative joint disease. 

In both datasets, the errors in the labels appear directly related to the weaknesses of the respective labelling methods. 

In the CXR14 data, the use of natural language processing on the reports is problematic because even if the process of label extraction is flawless, the reports themselves are often incomplete descriptions of the images. This hypothesis is supported by the large gap between the findings in Wang et al., which show that their labels are accurate reflections of the reports, and the visual appearance of the images. This discrepancy is understandable from a clinical perspective, as radiology reports are not simply an enumeration of image findings. Many findings are not included in radiology reports either because they are already known (for example, the classic report that only states “no change compared to previous”) or because the radiologist determined that the finding was not relevant to the referring clinician. This is a major concern because it suggests that label harvesting with natural language processing may never be able to accurately reflect the image findings on the films, and thus may never be able to produce high quality labels for training image analysis systems.

The CXR14 dataset also included examples of label disambiguation failure, with the majority of cases labelled “emphysema” actually showing evidence of subcutaneous emphysema, and of label schema failure, where the labels did not account for the clinically important stratification in the pneumothorax class. The majority of pneumothorax cases were already treated with chest tubes and often did not show radiographic evidence of ectopic pleural gas.

This latter point is important not only because untreated pneumothoraces are more clinically important to identify, but also because these labels are intended to train image analysis systems. While it is technically true that a patient with a chest tube in “has a pneumothorax”, if the majority of cases do not show any of the visual features associated with this pathology, the usefulness of the labels is highly suspect. What can we reasonably expect models to learn from these labels, other than the appearance of chest tubes?

The label schema of the CXR14 data also suffered from significant ambiguity, particularly related to the various labels for airspace opacities, to the point that it became almost impossible to design an acceptable way to relabel these classes. This was noted on review of the CXR14 labels themselves, as it was highly unclear why a case would be labelled as one class but not another. No common patterns were identified with exploratory analysis, suggesting that the differences between these labels may be mostly arbitrary.

The MURA labelling process was much more robust, because each radiologist at the point of care was asked to label each case as normal or abnormal. The issue with this labelling strategy arises because the definition of normal and abnormal cases was left up to each radiologist. Anecdotally, the Stanford team has suggested in private communication that many of the radiologists interpreted this to mean “normal for age”. This would be in keeping with the finding that degenerative joint disease in particular was under-reported in the labelling process, as presumably many radiologists may have decided that minor degenerative disease was within the expected range of normal for older patients. As the age of the patients is not provided with the dataset, this hypothesis cannot be explored further.

These weaknesses do not necessarily detract from the general usefulness of these datasets, but they do need to be understood if the models trained on the data are to function as expected. For example, a model trained on the MURA data should not be expected to detect hand or wrist osteoarthritis to any degree of accuracy. With this in mind, the biggest limitation of these datasets is not their label quality, but their documentation.

The supporting documents for these datasets do not adequately discuss these issues. In fact, the ChestXray14 paper and dataset FAQ explicitly state that “the text-mined disease labels are expected to have accuracy >90\%.“ Similarly, the MURA dataset paper presents an exploration of 100 abnormal cases, and states that the abnormalities include fractures, hardware, degenerative changes, and other miscellaneous findings. This gives the impression that the labels do cover degenerative disease, but is in fact an artefact of the process of only looking at the abnormal cases. If the team had reviewed the cases labelled normal, they would likely have discovered the presence of many cases of degenerative joint disease in this group as well.

In both circumstances, the original documentation is misleading. This raises an important question; “who is responsible for ensuring the quality of the data in public datasets?” The effort required to explore a dataset of this size is not negligible, and we may fear that expecting this level of analysis of teams who intend to release public datasets may dissuade them from producing these important resources.

It is also true that many end-users of this data are teams of computer scientists and engineers, who may not have easy access to the medical expertise required to understand the nuances of the data. Even worse, due to this lack of expertise, they may not even realise that the data could be flawed in the first place, particularly if they rely on test sets drawn from the same data.

On balance, the effort required to manually inspect a small subset of a dataset is fairly low compared to the effort required to build such a dataset, if performed a single time in a centralised manner (i.e., at the time of building the dataset). The team that builds the dataset is ideally suited to performing this analysis because they already understand the data generating process (for example, the MURA team already had anecdotal knowledge of how the labelling rule was being applied) and already have access to the medical expertise required for this assessment.

One way to partially mitigate the problems that users of the data may face is to produce a smaller second dataset purely for testing models trained on the original data, using a less flawed method, ideally involving expert visual review of cases. The MURA team has done this, using the majority vote of 3 board-certified radiologists to produce the labels for 207 randomly selected cases. Unfortunately, no analysis of the quality of the original labels for these 207 cases was presented in the MURA paper, nor was any subset analysis done on these test labels.

The CXR14 team did not provide a manually labelled test set. Independently, a team that published results on this dataset produced their own visual labels\textsuperscript{8}, and showed that the original labels significantly underperformed compared to other radiologists tested on the new labels (F1 score of 0.288 vs radiologist F1 scores of 0.35 to 0.44). Unlike the MURA test set, these CXR14 test labels are not publically available.
In both cases, the labelling rules used to produce these test sets were not explicitly stated.

While the use of a visually accurate set of labels for the test set does not solve many of the issues of the primary dataset, it does protect against the risk that the models will fail silently; the insidious risk that the model can reproduce the flawed labels but appear to be performing well because the test set is equally as flawed. 

There are a number of limitations to this analysis that should be acknowledged. First of all, the labels produced by LOR are not 100\% accurate. There will always be a significant amount of inter-observer variability, particularly when labels are ambiguous. This was compounded by the reduced image quality in each dataset. In particular, the reduction in the number of grey levels meant that many dense parts of the images became completely obscured, as if only a single window setting was available for review. For example it was regularly impossible to identify any retrocardiac pathology in the CXR14 dataset, because the heart appeared purely white.

These limitations were mitigated to some extent by being as permissive as possible when relabelling the CXR14 data, erring on the side of agreeing with the original labels.

It is also true that some of the labels in the CXR14 and MURA datasets are informed by information not available to the reviewer. This is probably particularly true in the case of airspace opacities, where a label of pneumonia or consolidation may be more likely to be used in a patient with a fever. However, in the context of producing labels for image analysis systems, it may actually be the case that a blind review of the images themselves is more worthwhile, as the presence of absence of image features alone is all that the models will be able to learn.

\section{Conclusion}
The disconnect between the dataset development and the usage of that data can lead to a variety of major problems in public datasets. The accuracy, meaning, and clinical relevance of the labels can be significantly impacted, particularly if the dataset development is not explained in detail and the labels produced are not thoroughly checked.

These problems can be mitigated by the application of expert visual review of the label classes, and by a thorough explanation of the development process, strengths, and weaknesses of the dataset. This exploration should include an analysis of the visual accuracy of the labels, as well as the identification of any clinically relevant subsets within each class. Ideally, this analysis and explanation will be part of the original release of the data, completed by the team producing the data to prevent duplication of these efforts, and a separate test set with visually accurate labels will be released alongside any large-scale public dataset.

\section*{References}

\small

[1]	Sun C, Shrivastava A, Singh S, Gupta A. Revisiting unreasonable effectiveness of data in deep learning era.  Computer Vision (ICCV), 2017 IEEE International Conference on; 2017: IEEE; 2017. p. 843-52.

[2]	Wang X, Peng Y, Lu L, Lu Z, Bagheri M, Summers RM. Chestx-ray8: Hospital-scale chest x-ray database and benchmarks on weakly-supervised classification and localization of common thorax diseases.  2017 IEEE Conference on Computer Vision and Pattern Recognition (CVPR); 2017: IEEE; 2017. p. 3462-71.

[3]	Rajpurkar P, Irvin J, Bagul A, et al. Mura dataset: Towards radiologist-level abnormality detection in musculoskeletal radiographs. arXiv preprint arXiv:171206957 2017.

[4]	Halabi SS, Prevedello LM, Kalpathy-Cramer J, et al. The RSNA Pediatric Bone Age Machine Learning Challenge. Radiology 2018: 180736.

[5]	Chilamkurthy S, Ghosh R, Tanamala S, et al. Deep learning algorithms for detection of critical findings in head CT scans: a retrospective study. The Lancet 2018; 392(10162): 2388-96.

[6]	Zbontar J, Knoll F, Sriram A, et al. fastmri: An open dataset and benchmarks for accelerated mri. arXiv preprint arXiv:181108839 2018.

[7]	Oakden-Rayner L. Exploring the ChestXray14 dataset: problems. 2018.

[8]	Rajpurkar P, Irvin J, Zhu K, et al. Chexnet: Radiologist-level pneumonia detection on chest x-rays with deep learning. arXiv preprint arXiv:171105225 2017.

\end{document}